 \documentclass[preprint2]{emulateapj}
 \usepackage{graphicx,makecell,multirow}

\shorttitle{A Search for AGN in the Reionization Era}
\shortauthors{Laporte et al.}

\begin{document}

\title{A spectroscopic search for AGN activity in the reionization era}

\author{Nicolas Laporte\altaffilmark{1},
Kimihiko Nakajima\altaffilmark{2},
Richard S. Ellis\altaffilmark{1,2},
Adi Zitrin\altaffilmark{3},
Daniel P. Stark\altaffilmark{4},
Ramesh Mainali\altaffilmark{4}, and
G. Roberts-Borsani\altaffilmark{1}
}

\altaffiltext{1}{Department of Physics and Astronomy, University College London, Gower Street, London WC1E 6BT, UK}
\altaffiltext{2}{European Southern Observatory (ESO), Karl-Schwarzschild-Strasse 2, 85748 Garching, Germany}
\altaffiltext{3}{Physics Department, Ben-Gurion University of the Negev, P.O. Box 653, Be'er-Sheva, 84105, Israel}
\altaffiltext{4}{Steward Observatory, University of Arizona, 933 N Cherry Ave, Tucson, AZ 85721, USA}


\begin{abstract}
The ubiquity of Lyman alpha (Ly$\alpha$) emission in a sample of four bright [O III]-strong 
star-forming galaxies with redshifts above 7 has led to the suggestion that such 
luminous sources represent a distinct population compared to their fainter, more 
numerous, counterparts. The presence of Ly$\alpha$ emission within the 
reionization era could indicate that these sources created early ionized bubbles due 
to their unusually strong radiation, possibly due to the presence of active 
galactic nuclei. To test this hypothesis we have secured long integration spectra 
with XSHOOTER on the VLT for three $z\simeq$7 sources selected to have similar 
luminosities and prominent excess fluxes in the IRAC 3.6 or 4.5$\mu$m band, usually 
attributed to strong [O III] emission. We secured additional spectroscopy 
for one of these galaxies at $z$=7.15 using MOSFIRE at the Keck telescope. For 
this, the most well-studied source in our sample with the strongest IRAC excess, 
we detect prominent nebular emission from He II and NV indicative of a non-thermal 
source. For the other two sources at $z$=6.81 and $z$=6.85, for which no previous 
optical/near infrared spectroscopy was initially available, Ly$\alpha$ is seen in one
and CIII] emission in the other. Although a modest sample, our results further support the 
hypothesis that the phenomenon of intense [O III]  emission is associated preferentially 
with sources lying in early ionized bubbles. However, even though one of our sources 
at $z$=7.15 clearly indicates the presence of non-thermal radiation, such ionized bubbles 
may not uniquely arise in this manner. We discuss the unique advantages of extending such 
challenging diagnostic studies  with JWST.
\end{abstract}




\section{Introduction} \label{sec:intro}
A fundamental challenge in supporting the now-popular claim that early star-forming 
galaxies are responsible for cosmic reionization (\citealt{2015ApJ...802L..19R}, \citealt{2016ARA&A..54..761S}) 
is the nature and strength of the ionizing radiation emerging from a typical source. 
To account for the optical depth of electron scattering seen by Planck \citep{2015A&A...576A.104P}, 
at least 10-20\% of the radiation produced by hot main sequence 
stars below the Lyman limit must escape scattering and absorption by clouds of 
neutral gas in the circumgalactic medium. Direct measures of this `escape fraction' 
are not yet possible beyond a redshift $z\simeq$3, and below which, following 
considerable observational effort, such a high fraction seems to be quite rare 
(\citealt{2013ApJ...765...47N}, \citealt{2015ApJ...810..107M}, \citealt{2016A&A...585A..51D}, \citealt{2016ApJ...825...41V}, 
\citealt{2017ApJ...841L..27R}, \citealt{2017MNRAS.468..389J}). Indirect methods based on tracing the 
extent of low ionization gas suggest the escape fraction may increase at higher 
redshift (e.g. \citealt{2013ApJ...779...52J}, \citealt{2016ApJ...831..152L}) but the validity of such 
methods remains unclear (\citealt{2016ApJ...831...38V}, \citealt{2016ApJ...828..108R}). 

The lack of evidence supporting the efficacy of star-forming galaxies as
producers of Lyman continuum radiation into the intergalactic medium (IGM)
has led some to suggest that a significant contribution of ionizing radiation
may emerge from non-thermal sources such as active galactic nuclei (AGN)
in the nuclei of more massive early galaxies. Assuming 100\% of such
non-thermal radiation can emerge into the IGM, \citet{2015ApJ...813L...8M}
demonstrate that, depending on the uncertain faint end portion of the
high redshift AGN luminosity function (\citealt{2011ApJ...728L..26G}, \citealt{2015A&A...578A..83G},
\citealt{2016arXiv160602719M}, \citealt{2017MNRAS.468.4691D}), a significant fraction of the late
ionizing contribution may arise in this manner. While it seems unlikely
early AGN can dominate the reionization process, such a contribution
could alleviate the requirement from star-forming galaxies. Of course,
given quasars with supermassive black holes are seen to redshifts
of at least $z\simeq$7, it seems reasonable to assume there are
earlier galaxies containing nuclear black holes.

Possible evidence in support of a contribution of ionizing
radiation from non-thermal sources follows the surprising
discovery of Lyman alpha (Ly$\alpha$) emission in all four $7.15<z<8.68$ sources
selected by \citet{2016ApJ...823..143R} (hereafter RBS) on the basis of
their high UV luminosity {\it and} the presence of intense
[O III] 5007 \AA\ emission as inferred from a prominent
excess signal in the IRAC 4.5$\mu$m band. This
visibility of Ly$\alpha$ contrasts markedly with the
results of more inclusive campaigns which targeted intrinsically
fainter sources which, at intermediate redshifts, normally have stronger line
emission (e.g. \citealt{2012ApJ...744..179S}, \citealt{2014ApJ...793..113P}).

Spectroscopic follow up of this unique sample (\citealt{2015ApJ...804L..30O}, \citealt{2015ApJ...810L..12Z}, \citealt{2017MNRAS.464..469S}) 
not only revealed ubiquitous Ly$\alpha$
emission at a time when the IGM is thought to be 60\% neutral,
but other UV nebular emission lines of high ionization potential
such as CIII] 1909 \AA\ . \citet{2017MNRAS.464..469S} suggest these [O III]-strong
luminous examples may have created early ionized bubbles
in the IGM thereby enabling Ly$\alpha$ photons to
escape. Although Stark et al proposed several different hypotheses
for the visibility of Ly$\alpha$ in these sources, the most intriguing posits
that these luminous sources harbor AGN.

The present paper is concerned with a detailed
a spectroscopic investigation of this hypothesis.
High ionization lines such as CIII] 1909 A, He II 1640,
CIV 1550 and NV 1240 \AA\ can be used as valuable
diagnostics of the underlying radiation field (e.g. \citealt{2016MNRAS.456.3354F}, \citealt{2016MNRAS.462.1757G},
\citealt{2017ApJ...836L..14M}). Although it is challenging
to detect these weaker lines with the necessary precision,
we have selected three bright sources with IRAC excesses
at $z\simeq$7 for a diagnostic study.

A plan of the paper follows. In Section \ref{sec:obs} we discuss
the selection of our 3 targets, one of which
is drawn from the earlier RBS sample. We also 
discuss the various spectroscopic campaigns.
The bulk of our data comes from deep exposures
with VLT's XSHOOTER which has the unique advantage
of simultaneous coverage of the bulk of the interesting
emission lines. We also discuss the reduction of the
data. In  Section \ref{sec:prop} we discuss the basic properties of our
sources based on SED-fitting and the known spectroscopic
redshifts. In Section \ref{sec:spectro}, we discuss the emission line spectra for each
of the sources in turn and, in Section \ref{sec:model} use photoionization
codes to test for the present of non-thermal radiation
in our sample. We summarize our results in 
Section \ref{sec:disc} in the context of future work.

Throughout this paper, a concordance cosmology is adopted, with $\Omega_{\Lambda}=0.7$, 
$\Omega_{m}=0.3$ and $H_{0}=70\ km\ s^{-1}\ Mpc^{-1}$. 
All magnitudes are given in the AB system \citep{1983ApJ...266..713O}. 

\section{Target Selection and Observations}\label{sec:obs}

The goal of the paper is to determine whether
the more luminous sources in the reionization era with
prominent [O III] excesses as detected with IRAC, reveal
evidence of AGN activity as determined from rest-frame UV
spectroscopy. 

The selection of an IRAC excess source is normally done
in addition to the now-standard Lyman break technique.
The detectability of an implied [O III] excess signal in either
the IRAC 3.6 or 4.5$\mu$m photometric band restricts 
the redshift range of targets. As discussed by \citet{2015ApJ...801..122S},
[O III] will be present in the 3.6$\mu$m band and display
an excess compared to 4.5$\mu$m only when H$\alpha$
does not lie in the latter band. This means a [O III] excess is optimally revealed for a
narrow redshift range $6.6<z<6.9$ corresponding to the
end of reionization. By contrast, a [O III] excess in the
4.5$\mu$m band does not suffer from an additional
line in the 3.6$\mu$m (except a likely weaker [O II] 3727 \AA\
for $z>7.6$).  In this respect, a 4.5$\mu$m excess arising
from [O III] is visible over $7<z<9$ \citep{2016ApJ...823..143R}.

However, there is a second consideration in the
spectroscopic follow-up of such candidates, namely
the visibility of the key diagnostic high ionization
potential metal lines \citep{2014MNRAS.445.3200S}. The most
valuable indications of AGN activity are the lines
of CIV 1550, He II 1640 and NV 1240 \AA\ \citep{2016MNRAS.456.3354F}.
With ground-based spectrographs, not all these lines
are well-placed 
between $z\simeq$ 7.3 and 8. 

The present sample was chosen on the basis of its
visibility from ESO's Very Large Telescope (VLT)
as well as maximizing the chance of detecting multiple
high ionization lines at $z\simeq$7. From the four
RBS targets discussed in \citet{2017MNRAS.464..469S}
we therefore selected the bright ($H_{160}$=25.1) target COS-zs7-1 with a
confirmed spectroscopic redshift of $z$=7.154
which has the largest 4.5$\mu$m excess in
the RBS list. In what follows we will refer to this
Y-band drop out as COSY.

To this, we added the two brightest sources drawn from the
list of 3.6$\mu$m excess objects published by \citet{2015ApJ...801..122S}, 
namely COS-3018555981($H_{160}$=24.9, $z_{photo}$=6.76)
and COS-2987030247 ($H_{160}$=24.8, $z_{photo}$=6.66).
As these are z-band dropouts, for convenience we refer below
to these as COSz1 and COSz2, respectively.

Although neither \citet{2015ApJ...801..122S} target was spectroscopically
confirmed at the time this project was conceived, we considered the
narrow redshift window for an excess in the 3.6$\mu$m
band to be a convincing indication. Subsequently (and fortunately),
both sources were spectroscopically confirmed via [C II] 158$\mu$m
detections at $z$= 6.85 and 6.81 respectively, with ALMA \citep{2017arXiv170604614S}. 

\begin{figure*}
\includegraphics[width=17cm]{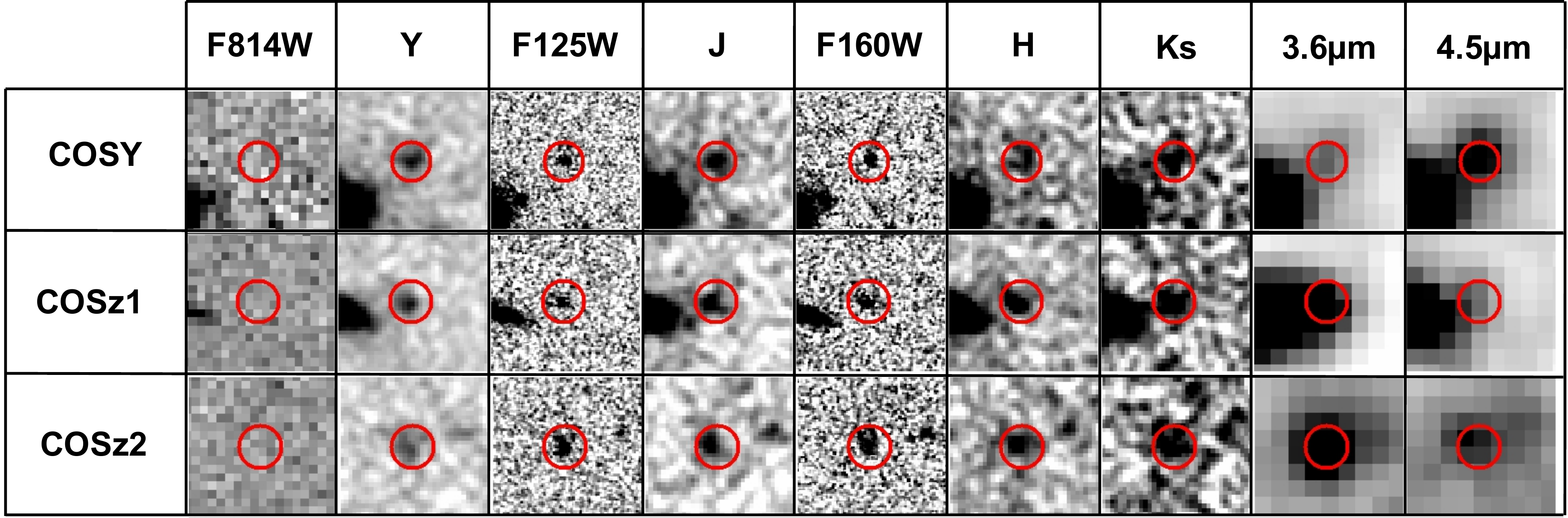}
\caption{\label{fig.stamps}Thumbnail images of the three $z\sim$7 targets observed with XSHOOTER/VLT. Each stamp is 5.5''$\times$5.5'', the position of each candidate is displayed by a red 0.8'' radius circle. }
\end{figure*}

All 3 targets lie in the COSMOS
field and thus there is excellent photometry from both
the CANDELS (\citealt{2011ApJS..197...35G}, \citealt{2011ApJS..197...36K}) 
and UltraVISTA \citep{2012A&A...544A.156M} surveys.
To aid our analyses we collated all the images containing
the 3 candidates using the public CANDELS \footnote{https://candels.ucolick.org/} (version 1) 
and UltraVISTA surveys\footnote{http://www.ultravista.org/} (version 3) catalogs. 
We added two images from the deep Spitzer-CANDELS  survey \citep{2015ApJS..218...33A} 
at 3.6 and 4.5 $\mu$m. 
Thumbnail images of our 3 targets in the various bands are shown in Figure \ref{fig.stamps} and
our derived photometry is in Table \ref{tab.photometry}.


\subsection{VLT Spectroscopy}

Our primary spectroscopic program was carried out with XSHOOTER/VLT \citep{2011A&A...536A.105V} in 
service mode (ID: 097.A-0043, PI: R. Ellis) between April 2016 and March 2017 at an average airmass of $\sim$1.2 and in 
good seeing conditions $\sim$0.7''.  Using blind offsets, observations were undertaken via A and B nodding positions 4 arcsec apart with
a 0.9 arcsec slit. Since the three XSHOOTER arms gather data independently, in order to maximise the key near-infrared 
exposures times, we adopted a unit 600s exposure, with 580s and 560s in the VIS and UVB arms respectively. The total
on source exposure time is 11h 20m (40.8 ksec) for COSz1 and COSz2, and 12h (43.2ksec) for COSY.

We reduced the XSHOOTER data using the ESO Reflex software (version 2.8 - \citealt{2013A&A...559A..96F})  together 
with the XSHOOTER pipeline (version 2.8.4). We first reduced all exposures on a given target with a master flat combining 
all flat exposures acquired during the same run. We also reduced all exposures with calibration data 
acquired during the same night, and combined all reduced exposures with the \textit{imcombine} task in IRAF. 
Tests demonstrated both resulting 2D-spectra are similar, and in the following we extract and analyse emission lines 
from spectra obtained with the first procedure.

\subsection{Keck Spectroscopy}

In a separate campaign (PI: A Zitrin), COSY was also studied in the J band using the MOSFIRE \citep{2012SPIE.8446E..0JM}
multi-slit spectrograph at the Keck observatory, on May 1 and 2, 2016.  A 0.7 arcsec wide slit was placed on COSY and the 
observations were carried out in 120s unit exposures with an AB dithering pattern of $\pm$1.5 arcsec along the slit.  Among 
other objects, another slit in the same mask was placed on a nearby star to monitor possible drifts and changes in seeing and transparency. 
Exposure times of 1.8 and 2.6 hours on the first and second night, respectively, were obtained, for a total integration time of 4.4h. 
The average airmass during the observations was $\sim1.2$, the average seeing 0.7 arcsec, and the conditions clear. 


The MOSFIRE data were reduced using the standard MOSFIRE reduction pipeline \footnote{http://www2.keck.hawaii.edu/inst/mosfire/drp.html}, 
which includes flat-fielding, wavelength calibration, background subtraction, and combining individual exposures for each slit on the mask. 
The output yields a reduced 2D spectrum per slit per night, along with its error and S/N . The combined 2D spectrum was obtained by 
inverse-variance weighting the resulting 2D spectra from the two nights. 

Finally, on the same campaign, a further object from the list of \citet{2015ApJ...801..122S}, namely COS-1318939512. ($H_{160}=25.0\pm0.1$, $z_{photo}=6.75^{+0.09}_{-0.08}$),  
was observed with the same integration time, but no significant line emission was detected. 
Since the spectroscopic redshift of this target is not yet clear, it is not included in this analysis.

\begin{table*}
\small
\centering
\caption{\label{tab.photometry} Photometry of the three selected targets. Non-detections are at 2$\sigma$ in a 0.4'' radius aperture at the object's position.}
\begin{tabular}{|c|ccccccccc|}
\hline
ID 	&	F814W 	&	Y	&	F125W	&	J	&	F160W	&	H	&	Ks	&	3.6$\mu$m	&	4.5$\mu$m \\
\hline
COS$Y$	&	$>$ 30.1	&	25.09$\pm$0.14	&	25.25$\pm$0.13	&	25.11$\pm$0.17	&	25.32$\pm$0.17	&	25.47$\pm$0.30	& 	25.35$\pm$0.37	& 25.09$\pm$0.59	&	23.92$\pm$0.32 \\	
COS$z1$	&	$>$29.4	&	25.43$\pm$0.19	&	25.35$\pm$0.15	&	25.31$\pm$0.21	&	25.09$\pm$0.14	&	25.18$\pm$0.23	&	25.12$\pm$0.30	&	23.90$\pm$0.30	& 25.20$\pm$0.60 \\
COS$z2$&	$>$29.4	&	25.53 $\pm$0.21&	24.85$\pm$0.1	&	25.43$\pm$0.23	&	25.38$\pm$0.18	&	25.30$\pm$0.26	&	25.06$\pm$0.28	&	24.11$\pm$0.45	&	24.39$\pm$0.48 \\
COS$z2^{\star}$	&	$>$29.4	&	26.76 $\pm$0.37&	26.89$\pm$0.32	&	26.40$\pm$0.32	&	25.79$\pm$0.13	&	25.64$\pm$0.21	&	26.01$\pm$0.38	&	-	&	- \\
\hline
\end{tabular}
\end{table*}

\section{Physical Properties }\label{sec:prop}

Before we discuss the spectroscopic diagnostics (Section \ref{sec:spectro}), we review the physical properties of our three [O III]-strong sources so as to place them in the context of other sources being targeted in the reionization era.

\begin{figure}
\centering
\includegraphics[width=8cm]{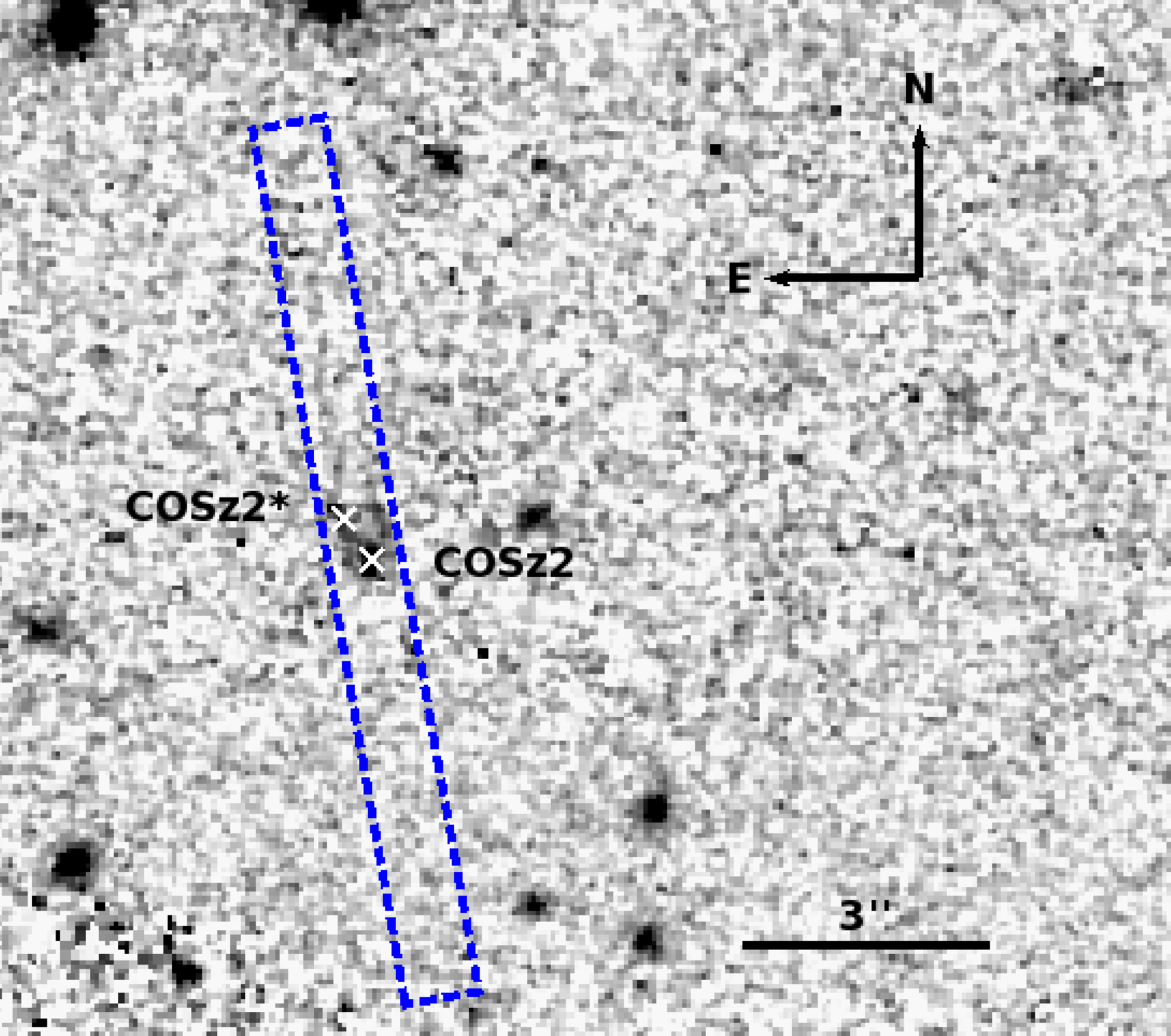}
\caption{\label{fig.cosz2} Detailed HST F160W image of the target COS$z2$ target indicating the presence of the companion COS$z2*$ 0.7 arcsec to the NE whose photometric and spectroscopic properties suggest it is a foreground object (Table \ref{tab.compar_photoz}). The centroid of each object is displayed by the white cross, the position of the XSHOOTER slit is displayed by the dashed blue box.}
\end{figure}

Firstly, given we have accurate spectroscopic redshifts, both from our earlier work (Stark et al 2017) and the ALMA data (Smit et al 2017), it is instructive to examine the accuracy of the earlier photometric redshifts derived from the extensive photometry in the COSMOS field (Table 1). This is additionally important since, in the case of COS$z2$, \citet{2015ApJ...801..122S} identified a nearby object 0.7 arcsec to the NE on the HST image (Figure ~ \ref{fig.cosz2}) which may confuse our measurements. We used strong SExtractor deblending parameters (DEBLEND\_NTHRESH 62 and DEBLEND\_MINCOUNT 0.0000001) to extract individual fluxes for COS$z2$ and its companion (which we refer to as COSz2$^{\star}$). We used both the {\tt Hyperz} \citep{2000A&A...363..476B} and {\tt BPZ} \citep{2011ascl.soft08011B} codes for this investigation of photometric redshift precision and, as we are combining photometry from different instruments, we applied the method described in \citet{2007ApJ...670..928B} taking as reference the SExtractor MAG\_AUTO in F160W and Ks for the HST and UltraVISTA datasets respectively. 

For  \textit{Hyperz} we adopted 377 templates from \citet{1997A&A...326..950F}, \citet{1998ApJ...509..103S}, \citet{2003MNRAS.344.1000B}, \citet{1980ApJS...43..393C}, \citet{1996ApJ...467...38K}, \citet{2007ApJ...663...81P}, \citet{2010A&A...514A..67M} and \citet{2001ApJ...556..562C} plus emission lines templates as described in \citet{2014A&A...563A..81D}.  We allowed a redshift range between 0.0 and 10.0 with extinctions $A_V$ from 0.0 to 3.0 mag. The three targets are well fitted with reduced $\chi^2 <1.7$ and redshift probability distributions with $z_{photo}>6$. For the companion COSz2$^{\star}$, excluding the IRAC photometry we find $z_{photo}$ = 2.16$^{+0.65}_{-0.15}$ with a reduced $\chi^2$ of 1.21.  \textit{BPZ} uses templates derived from 8 SEDs (\citealt{1980ApJS...43..393C}, \citealt{1996ApJ...467...38K}, \citealt{1997AJ....113....1S}) and we defined the prior in luminosity using the $J-$ band magnitude. Again, all candidates have $z>6$ solutions with a reduced $\chi^2 <$ 1.5.  For COSz2$^{\star}$, we again found a low-$z$ solution at $z_{photo}$ = $1.50$ but with less certainty. Table \ref{tab.compar_photoz} summarises the results.

\begin{figure*}
\centering
\includegraphics[width=15cm]{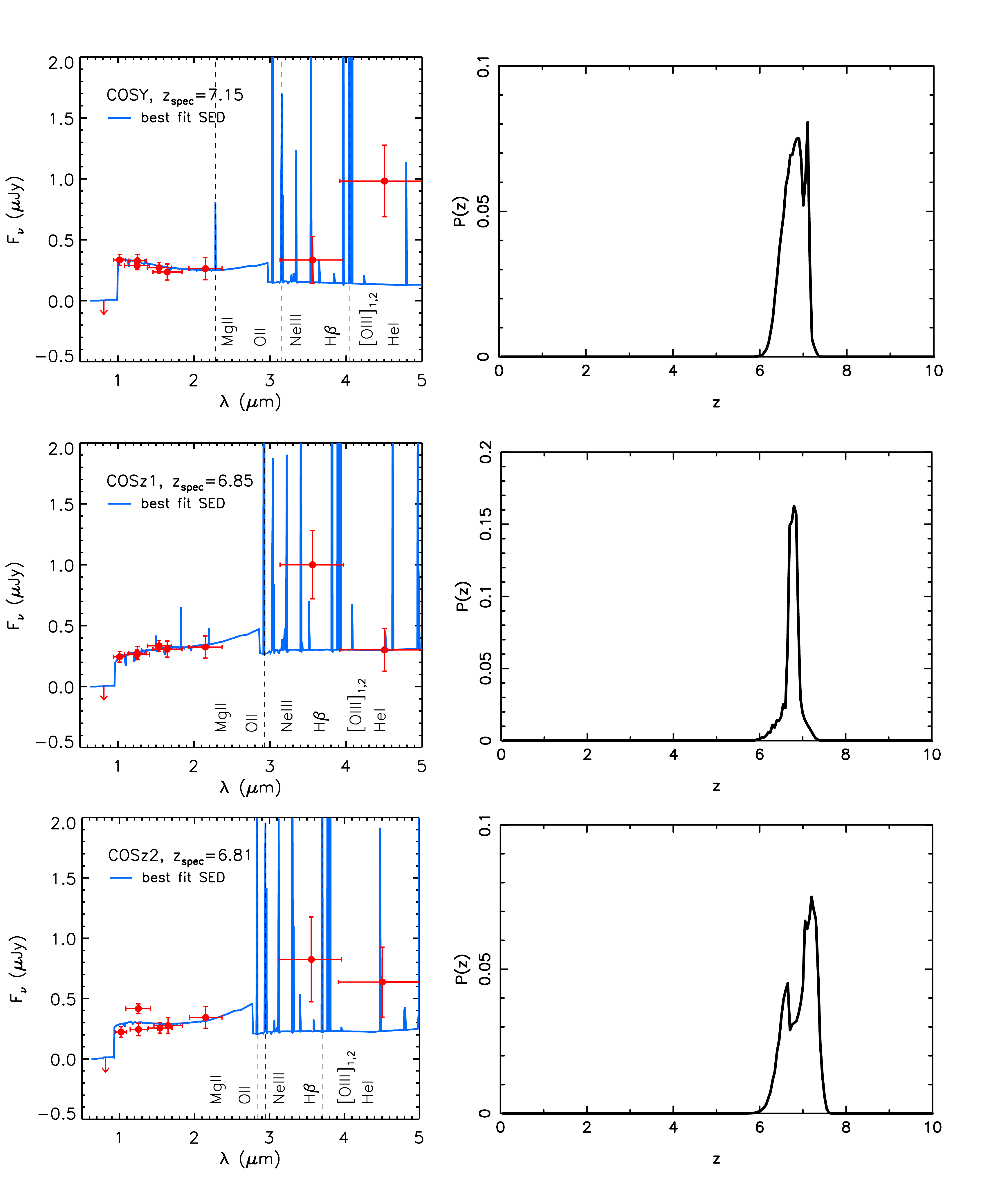}
\caption{(Left) Spectral Energy Distribution of the three targets. (Right) Photometric redshift likelihood functions (see text for details).}
\end{figure*}

\begin{table*}
\centering
\caption{\label{tab.compar_photoz} Photometric redshift estimates obtained using different SED-fitting approaches : \textit{Hyperz} uses a $\chi^2$ minimisation method and \textit{BPZ} uses a Bayesian procedure.}
\begin{tabular}{|c|ccc|ccc|c|}
\hline
\multirow{2}*{ID} 	  	&	\multicolumn{3}{c|}{\textit{Hyperz}}  & \multicolumn{3}{c|}{\textit{BPZ}}  & \multirow{2}*{$z_{spec}$} 	  \\ 
 \cline{2-7}&	 $z_{phot}$	& $\chi^2$		&	1$\sigma$		&	$z_{phot}$	& $\chi^2$		&	1$\sigma$  & \\
\hline
COS$Y$		   &	7.06	& 0.67	& 6.21 - 7.21 & 6.95 	&	1.12	& 6.17 - 7.73  & 7.149 \\
COS$z1$		   &	6.84	& 1.26	& 6.25 - 7.21 &	6.90	&	0.74	& 6.12 - 7.68  & 6.854 \\
COS$z2$		   & 	6.61	& 1.71	& 6.08 - 7.46 & 6.88 &	1.43	& 6.11 - 7.65 &  6.816\\
COS$z2^{\star}$ & 	2.16 & 1.21	& 2.01 - 2.81 & 1.50	&	0.76	& 1.14 - 8.32 &  2.11\\
\hline
\end{tabular}
\end{table*}

Adopting the spectroscopic redshifts, we next use the SEDs to determine their important physical properties, namely stellar masses, star formation rates (SFRs) and UV luminosities, $M_{UV}$. This allows us to consider them in the context of other sources being surveyed in the reionization era. We begin by using the \textit{MAGPHYS} \citep{2008MNRAS.388.1595D} code. Since this code does not take into account the possibility of contamination by nebular emission, we ignore the photometry in the relevant IRAC band. We find that all targets have stellar masses ranging from 0.19 to 1.1$\times$ 10$^{10}$ M$_{\odot}$, SFRs ranging from 20 to 33 M$_{\odot}$/yr and dust contents with $A_V$ ranging from 0.3 to 0.9 mag. 

\begin{figure}
\centering
\includegraphics[width=10cm]{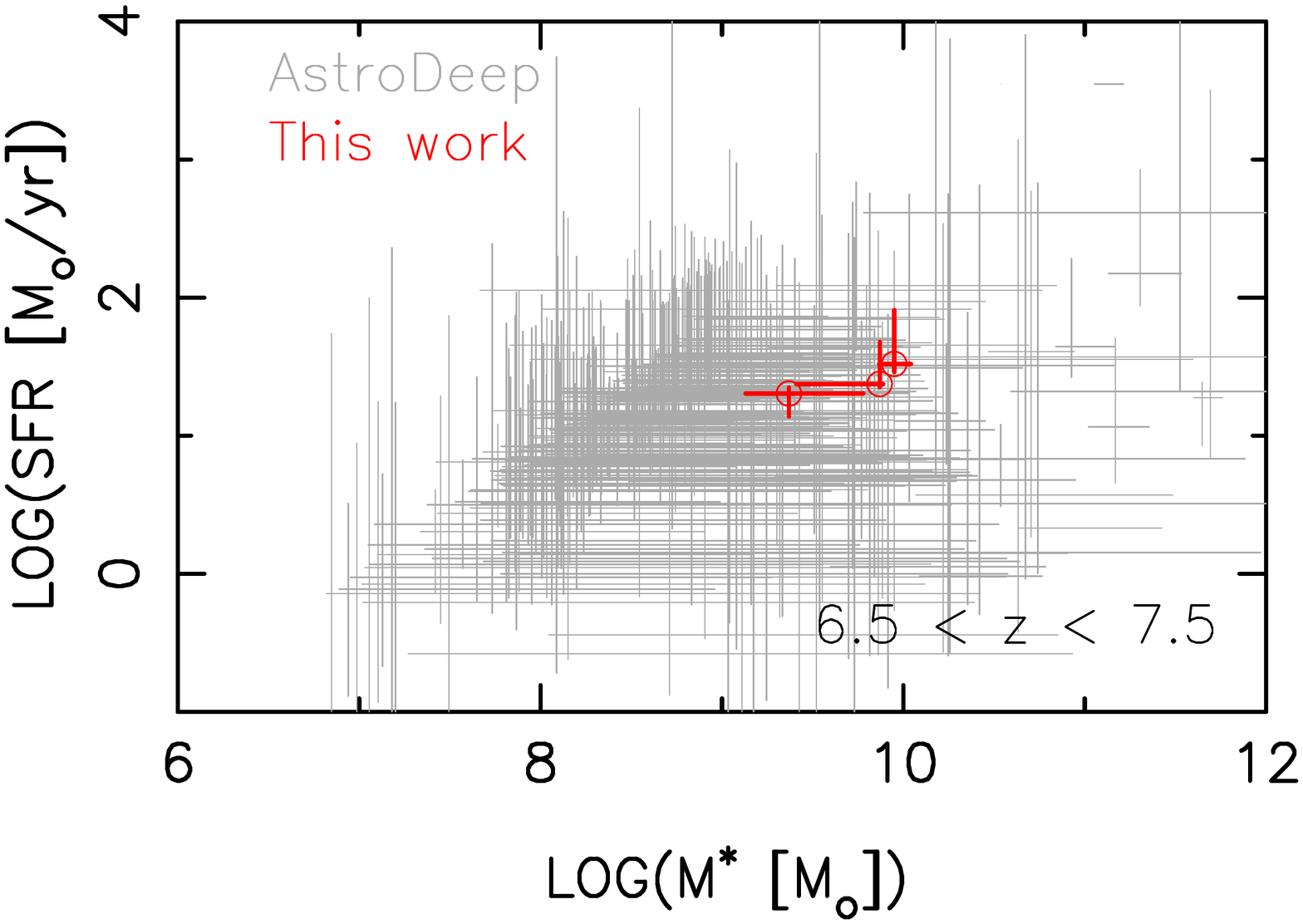} \\
\caption{\label{fig:SFR} The star formation rate and stellar mass of the three targets (in red) are compared with those of well-fitted $z\sim$7 objects in the AstroDeep catalogs (in grey) for four of the Frontier Fields ( \citealt{2017arXiv170603790D}, \citealt{2016A&A...590A..30M} and \citealt{2016A&A...590A..31C} ). Clearly the targets represent the most massive and actively star-forming sources at these redshifts. }
\end{figure}

To consider more carefully the influence of the inferred [OIII] emission line in the IRAC photometry on the derived stellar mass, we next use \textit{Hyperz} with the Starburst99 library \citep{1999ApJS..123....3L} again adopting the spectroscopic redshift. Assuming an age prior of $>$ 10Myr, we now find the stellar masses range from 2.3 to 8.9$\times$10$^{9}$  M$_{\odot}$. In order to place our targets in context with the general population of objects at $z\sim$7, we make use of the AstroDeep catalogues publicly available for 4 of the 6 Frontier Fields (\citealt{2017arXiv170603790D}, \citealt{2016A&A...590A..30M} and \citealt{2016A&A...590A..31C}). We compare our sources with all sources from this catalogues with 6.5 $< z_{phot}  <$ 7.5, which have been well fitted ( 0.5 $< \chi^2_{reduced}<$ 2.0) and with a narrow redshift probability distribution ($\Delta z<$1.0). In Figure \ref{fig:SFR} we plot the SFRs and stellar masses of our 3 targets comparing them with those in the AstroDeep population. As seen, we are probing the properties of the most massive objects at $z\sim$7.  We also estimated physical sizes using the half light radius measured by SExtractor on the F160W HST image (following the method of \citealt{2010ApJ...709L..21O}). These range from 0.3 to 1.0 kpc consistent with results by \citet{2010ApJ...709L..21O},  \citet{2015ApJ...804..103K}, and \citet{2016ApJ...820...98L}. Therefore they display a specific Star Formation Rate of 3-9$\times$10$^{-9}$ yr$^{-1}$ which is consistent with those previously published by \citet{2013ApJ...763..129S}, \citet{2014MNRAS.444.2960D} and \citet{2015A&A...577A.112L}. We also compute the UV slope $\beta$ following the precepts of \citet{2014ApJ...793..115B}.  There is quite a variation within our sample with COS$z1$, and to a lesser extent, COS$z2$, significantly redder than COSY and the general population.

We summarize these physical properties in Table \ref{tab.properties} adopting the \textit{Hyperz} derivations for the stellar mass and photometric redshift, and \textit{MAGPHYS} results for the SFR and reddening.

\begin{table*}
\centering
\caption{\label{tab.properties} Physical properties deduced from the SEDs and spectroscopic redshifts.}
\begin{tabular}{|c|cc|c|c|cc|c|cc|}
\hline
\multirow{2}*{ID}	&	RA  		& DEC   	& $z_{spec}$	& M$_{\star}$	&  SFR 			& 	$A_v$	&	$r_{1/2}$ 	&	M$_{UV}$		&	UV slope\\ 
				&	[J2000]	& [J2000]	&			& $\times$ 10$^{9}$ M$_{\odot}$	& M$_{\odot}$/yr	&	[mag] 	& [kpc] 	&	&	 \\
\hline

COS$Y$			& 150.09904	& 2.3436043	& 7.149	& 2.34 $^{+3.69}_{-0.99}$		& 20.2  $^{+2.2}_{-6.4}$	& 0.3 $^{+0.3}_{-0.1}$ 	& 0.33 $\pm$0.03	&	-21.8	 $\pm$ 0.2	& -2.33 $\pm$ 0.03	 \\
COS$z1$			& 150.12575	& 2.26661	& 6.854	& 7.41  $^{+0.35}_{-4.84}$	& 23.7 $^{+24.2}_{-1.3}$	& 0.8 $^{+0.1}_{-0.4}$	& 0.79 $\pm$0.21 	&	-21.6 $\pm$ 0.2		& -1.18 $\pm$ 0.19	\\
COS$z2$			& 150.12444	& 2.21729	& 6.816	& 8.91  $^{+1.56}_{-2.15}$	& 33.1 $^{+48.2}_{-4.2}$	& 0.3 $^{+0.4}_{-0.1}$	& 0.96$\pm$0.20	&	-22.1 $\pm$ 0.1		& -1.72 $\pm$ 0.43	 \\

\hline
\end{tabular}
\end{table*}

\section{Emission Line Detections}\label{sec:spectro}

\begin{figure*}
\centering
\includegraphics[width=19cm]{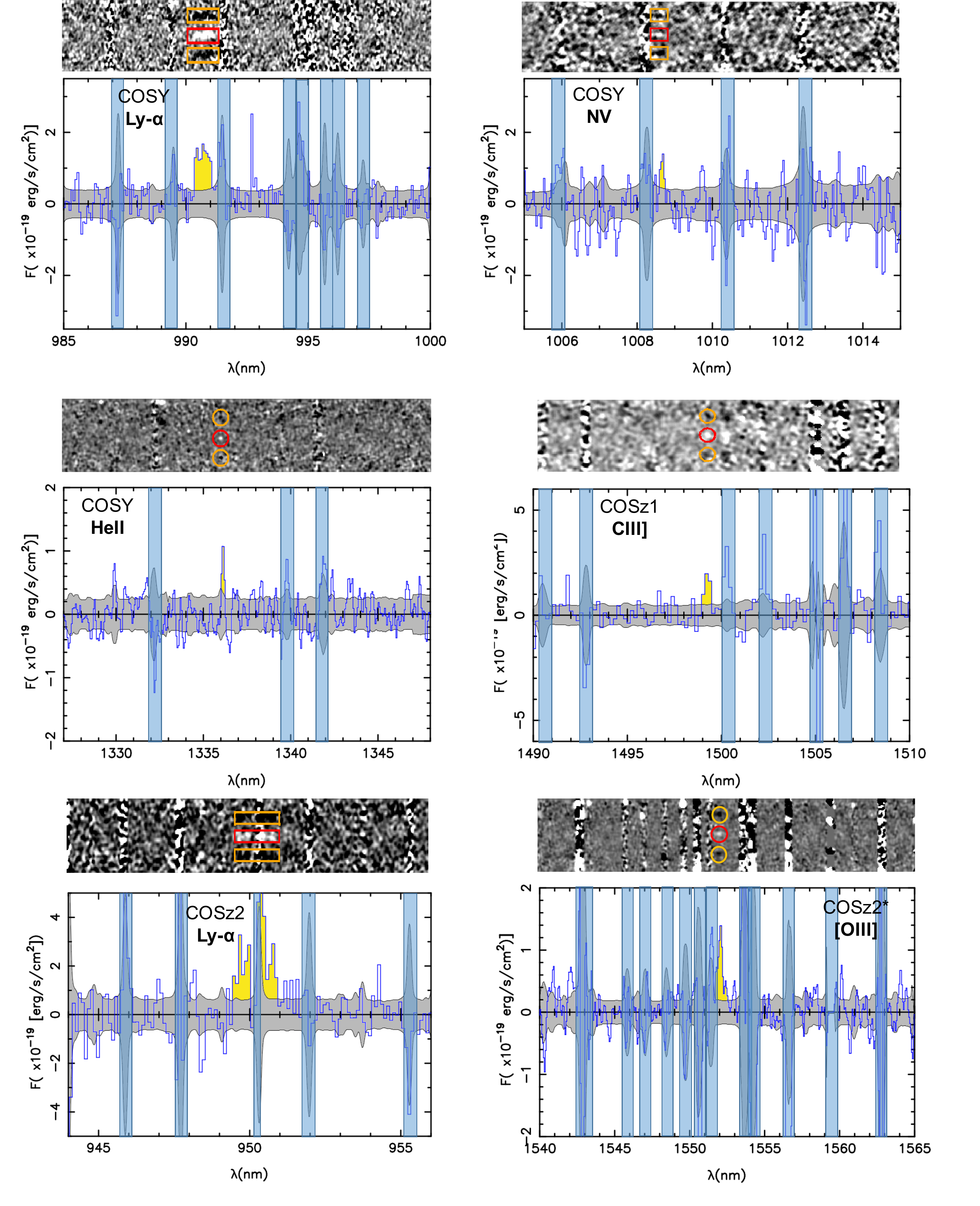} \\
\caption{\label{fig.lines} For each panel, we show the XSHOOTER 2D-spectrum (upper) revealing the two negative counterparts with the central positive and (lower) the extracted 1D-spectrum (blue) with the 1$\sigma$ rms (gray). \textit{Top-left} : Ly-$\alpha$ emission line detected in COSY, \textit{Top-right}: NV emission in COSY,  \textit{Center-left} : HeII emission in COSY, \textit{Center-right} : [CIII] in COS$z1$, \textit{Bottom-left} : Ly-$\alpha$ in COS$z2$,  \textit{Bottom-right} : [OIII] in COS$z2^{\star}$.}
\end{figure*}

All 3 targets are massive sources with intense star-formation rates so we now examine the nature of their radiation fields as well as to explore whether, as was the case for the four RBS [O III]-excess sources, the newly-studied sources also show prominent Ly$\alpha$ emission. We discuss the emission line content of the spectra of each target in turn based both on a visual inspection of the 2D spectra, and the 1D extracted spectrum at the relevant target position within the slit. Recognizing that the reliable detection of faint diagnostic lines is challenging even with these long integration times, we verified the credibility of weak lines by examining the two independent half-exposures and, in the specific case of COSY, comparing the XSHOOTER and MOSFIRE spectra where they overlap in wavelength. In the following discussion, errors on line fluxes and upper limits were determined by considering the signal in adjacent regions using a rectangular aperture set by the width of typical nebular lines (12 pixels for XSHOOTER) and the spatial extent set by the seeing (3 pixels). The relevant line detections are collated and displayed in Figure \ref{fig.lines}.

\subsection{COSY}

\begin{figure}
\centering
\includegraphics[width=9cm]{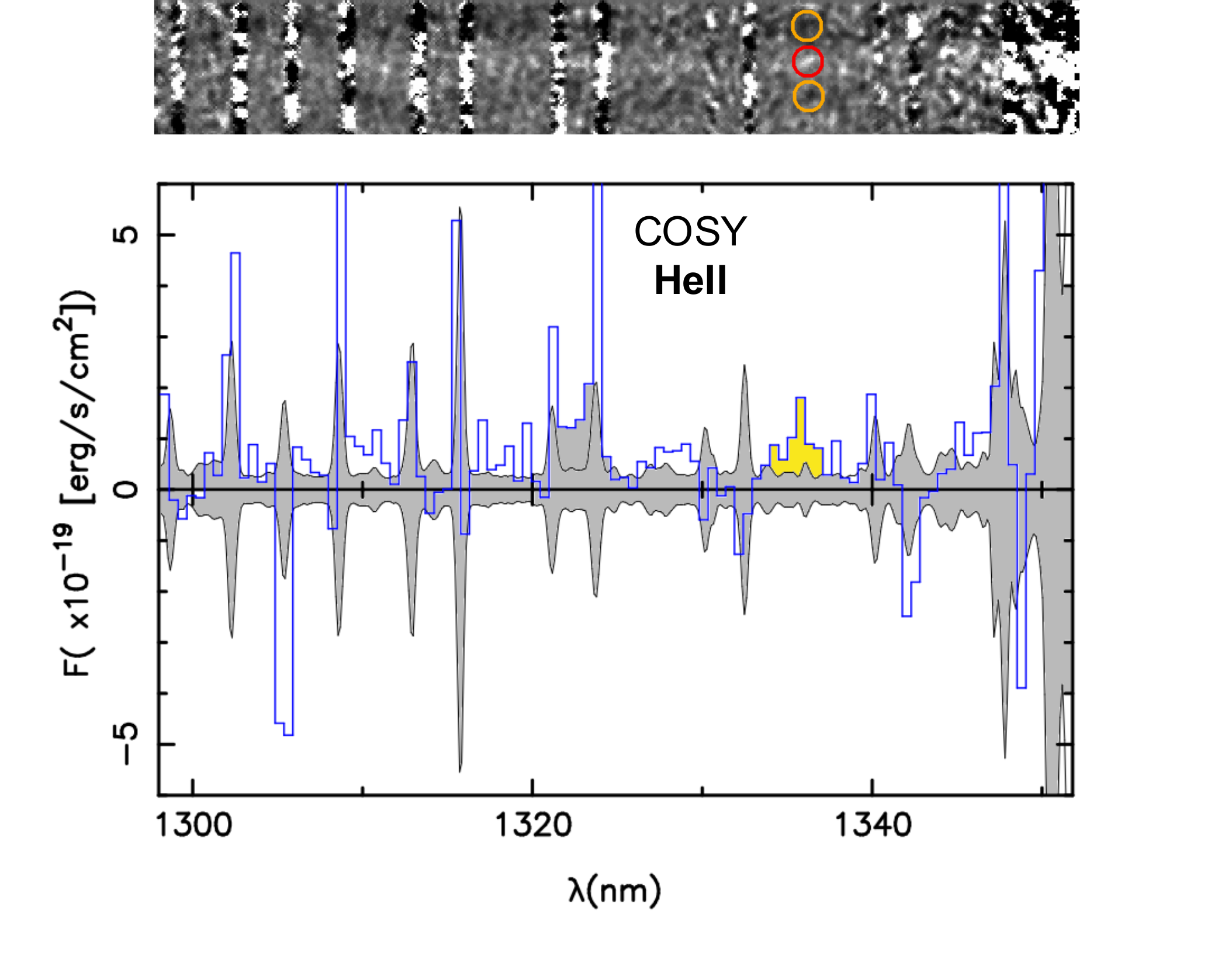} \\
\caption{\label{fig.HeII_MOSFIRE} As Figure \ref{fig.lines} for the HeII emission line detected in the COSY spectrum obtained with MOSFIRE. }
\end{figure}

\citet{2017MNRAS.464..469S} already detected prominent Lyman-$\alpha$ emission at $z=$7.15 in this target on the basis of a 4 hour exposure with MOSFIRE. Furthermore, deep ALMA observations provide an additional detection of [CII] 158$\mu$m \citep{2016ApJ...829L..11P}. With the deeper XSHOOTER data we recover Ly$\alpha$ at 9907.2 \AA\ with significantly improved signal to noise and note it is reasonably broad. Additionally, we detect two further emission lines at 10 086.4 \AA\  and 13 360.3 \AA\  which we identify as NV 1240 \AA\ and HeII 1640\AA\  respectively. Both new lines are significant at $\approx$5$\sigma$ with rest-frame equivalent widths (EWs) of 3.2$^{+0.8}_{-0.7}$\AA\ and 2.8$^{+1.3}_{-0.9}$\AA\ respectively for NV and HeII. A potential detection at 9927 \AA \ is rejected due to the absence of negative counterparts. We estimate a Ly-$\alpha$ EW = 27.5$^{+3.8}_{-3.6}$ which is consistent with previous findings by \citet{2017MNRAS.464..469S} and a velocity offset of $\Delta{\nu}_{Ly\alpha}$=286.6 km/s, based on a previous [CII] detection from \citet{2016ApJ...829L..11P}. Such an offset is similar to those observed in $z\geq$ 6.5 spectroscopically-confirmed galaxies.  The limited overlap in wavelength between the independent MOSFIRE spectrum and that of XSHOOTER allows us to confirm the HeII 1640\AA\ emission at $\lambda$=13 357 \AA\ with a flux of (1.20$\pm$0.45)$\times$10$^{-18}$ erg/s/cm$^{-2}$, consistent with the XSHOOTER data (Figure \ref{fig.HeII_MOSFIRE}).

\subsection{COS$z1$}

Only one emission line at $\lambda$ = 14 992\AA\ is seen in this source over the full wavelength range covered by XSHOOTER . The line appears to be robust with a $\approx$4 $\sigma$ significance ($f_{CIII}$=1.33$\pm$0.31 cgs, rest-frame EW=4.0$^{+2.2}_{-1.5}$\AA\ ). Adopting the ALMA redshift from \citet{2017arXiv170604614S}, the likely identification is CIII]1909\AA\ at $z$=6.854. In this case, normally we would expect to see the 1907\AA\ companion in the doublet at 14 977\AA\ since the night sky spectrum is relatively clear at this wavelength. The 1907/1909 \AA\ flux ratio is typically $\simeq$1.0-1.5 at intermediate redshifts.  Our non-detection of CIII] 1907 implies a flux ratio of $<0.6$ at 3$\sigma$. This could imply a very high gas density. Upper limits are tabulated for the other diagnostic lines in Table \ref{tab.fluxes}; that for Ly$\alpha$ is particularly stringent.

\subsection{COS$z2$}

We detect two reasonably convincing emission lines in the COS$z2$ XSHOOTER spectrum at $\lambda$=9502.0 \AA\ (S/N$\sim$ 3) and $\lambda$= 15 520 \AA\ (S/N$\sim$7.0). However, we cannot find any association at the same redshift for these two lines, and therefore, noting the double structure of COS$z2$ discussed in Section \ref{sec:prop}, examined whether the lines may belong to two independent sources. Assuming that the bluer line is Lyman-$\alpha$, the redshift of COS$z2$ would be $z$=6.816 with a rest-frame EW= 16.2 $^{+5.2}_{-5.5}$ consistent with a velocity offset $\Delta_{\nu}$=325.6 km s$^{-1}$ comparable to previous findings given the ALMA redshift. As in the case of COSY, Ly$\alpha$ appears unusually broad compared to other detections at this redshift. Unfortunately, some of the other diagnostic lines (such as HeII, CIII] ) would be hard to detect at this redshift as they would lie close to, or be obscured by, a night sky line . It seems likely that the second line is associated with the nearby companion,COS$z2^{\ast}$. If this line is [OIII] 5007 \AA\  the redshift would be be $z$=2.099 and its weaker 4959 \AA\ component would fall under a sky line. Although the photometric redshift is uncertain ($z_{phot}$=2.16$^{+0.65}_{-0.15}$), the identification is consistent. 

\begin{table*}[h!]
\centering
\caption{\label{tab.fluxes} Extracted fluxes and upper limits for the various emission lines with errors determined by considering the signals in several apertures at similar wavelengths (see text for details). Individual detection uncertainties are 1$\sigma$ and non-detection are quoted at the 3 $\sigma$ level. Velocity offsets are computed from the [CII] detections by ALMA (\citealt{2016ApJ...829L..11P} and \citealt{2017arXiv170604614S})}
\begin{tabular}{|c|cccc|r|}
\hline
Object 				& 	Line				& $\lambda_{rest frame}$ 		& $\lambda_{obs}$ 	&	flux	 & $\Delta_{\nu}$ \\  
					&					& [\AA\ ]					& [\AA\ ]			&	$\times 10^{-18}$ erg/s/cm$^2$  	& km/s      \\ 	\hline
\multirow{5}*{COSY}  	&	Lyman-$\alpha$	& 1215.7					& 9 907.2			& 	22.9$\pm$3.0					& 286.6	 \\
					&      NV				& 1238.8 					& 10 086.4 		& 	2.58$\pm$0.44 					& 17.	4	\\
					&	CIV				& 1548.2 					& -				&      $<$2.70						&  \\
					& 	HeII				& 1640.0 					& 13 360.3		& 	1.26$\pm$0.29 					&  181.4 \\ 
					&	CIII]				& 1906.7  /1908.7 			& -				&	$<$ 0.92	/ 	$<$0.83			&	\\ \hline	
\multirow{5}*{COSz1}	&	Lyman-$\alpha$	& 1215.7					& -				& 	$<$ 0.96						& \\				
					&      NV				& 1238.8 					& - 				& 	$<$ 1.62						& \\
					&	CIV				& 1548.2 					& -				&      $<$ 2.22				 		& \\
					& 	HeII				& 1640.0 					& -				& 	$<$ 2.88						&\\ 
					&	CIII]				& 1906.7  /1908.7 			& - / 14 992.0		&	$<$ 1.18 / 1.33$\pm$0.31 		& 21. 4\\ \hline

\multirow{5}*{COSz2}	&	Lyman-$\alpha$	& 1215.7					& 9502.0			& 	9.35$\pm$3.7					 & 325.6 \\				
					&      NV				& 1238.8 					& - 				& 	$<$	1.53						 & \\
					&	CIV				& 1548.2 					& -				&      $<$ 	2.49					 	 & \\
					& 	HeII				& 1640.0 					& -				&      \textit{Sky-line}					 & \\ 
					&	CIII]				& 1906.7  /1908.7 			& -				&	\textit{Sky-line} 	/ $<$ 1.57		 	 & \\ \hline
 
\hline
\end{tabular}
\end{table*}

\section{Photoionization Modeling}\label{sec:model}

In summary, our spectra reveal Ly$\alpha$ emission for two of our 3 IRAC excess targets weakening slightly the ubiquity of emission that was so striking in the earlier RBS sample. However, we see evidence for non-thermal radiation field in one source and possibly a harder radiation field than typical at lower redshift in a second. The spectrum of COSY is particularly intriguing with convincing detections of NV and He II. COS$z1$ also shows a convincing CIII] emission as was found by \citet{2017MNRAS.464..469S} for the $z$=7.73 source EGS-z8-1. Other than an indication that Ly$\alpha$ is unusually broad, there is no evidence of a hard radiation field for COS$z2$; although strong He II cannot be detected due to the coincidence with a sky line, there is no evidence of CIV or NV emission. We now examine the consequences of these emission line ratios, focusing on COSY and COS$z1$, using a suite of photoionization models discussed more fully in Nakajima et al (2017, in prep). We briefly summarize the details of these simulations below.

Our photoionization models are based on {\it Cloudy} (version 13.03; \citealt{1998PASP..110..761F,2013RMxAA..49..137F}) and assume constant-density gas clouds with a plane-parallel geometry. Dust physics and elemental depletion factors follow the analyses of \citet{2006ApJS..167..177D} and \citet{2011Sci...333.1128N}. All elements except nitrogen, carbon and oxygen, are taken to be primary nucleosynthetic products. We use the precepts given by \citet{2006ApJS..167..177D} and \citet{2012MNRAS.426.2630L}, respectively, to account for their secondary products. For helium, we adopt the form given by \citet{2006ApJS..167..177D}. The models are constructed for varying ISM properties of metallicity ($Z$), ionization parameter ($\log U$), and electron density. 

To test the AGN hypothesis, we contrast the predictions for ionizing radiation fields from star-forming galaxies generated by stellar population synthesis codes (including both single and binary stellar evolution) and AGN with a range of power law indices. An important caveat is that the models do not consider the effects of shocks.  Although the effect of shocks on the integrated UV spectra of star-forming galaxies is expected to be modest, \citet{2016ApJ...833..136J} have shown how CIII] emission can be enhanced by shocks with low velocities and strong magnetic fields, leading to spectral signatures mimicking those of AGN.

For the radiation field from star-forming galaxies, we adopt the population synthesis code BPASS (v2; \citealt{2015MNRAS.446.3911S}). We use publicly available BPASS SEDs for a Kroupa IMF under the assumption of a continuous star-formation history at an age of 50Myr. Stellar metallicities are matched to their gas-phase equivalents. Both single and binary star populations are considered. 

For the AGN, we consider a narrow-line region (NLR) surrounding an AGN ionizing radiation field characterized by a power law. Our AGN models are generated by the {\it Cloudy} ``AGN'' continuum command with the default parameters except for the power-law energy slope between the optical and X-ray bands, $\alpha_{ox}$ \citep{1981ApJ...245..357Z}. This parameter corresponds to the power-law index $\alpha$, where $f_{\nu}\propto \nu^{\alpha}$, determined in the range of a few to a few thousand eV. The AGN models are truncated at a neutral column density of $N(HI) =10^{21}$\,cm$^{-2}$, in accordance with the NLR models of \citet{2013ApJ...774..100K}. For the AGN modes we assume dust-free gas clouds and ignore the depletion of elements onto dust grains, consistent with models that reproduce observations of high-redshift radio galaxies, type-II QSOs, and local type-II AGNs (e.g.,\citealt{2006A&A...447..157N}).

\begin{figure}
\vspace{-0.5cm}
\includegraphics[width=7cm, angle=-90]{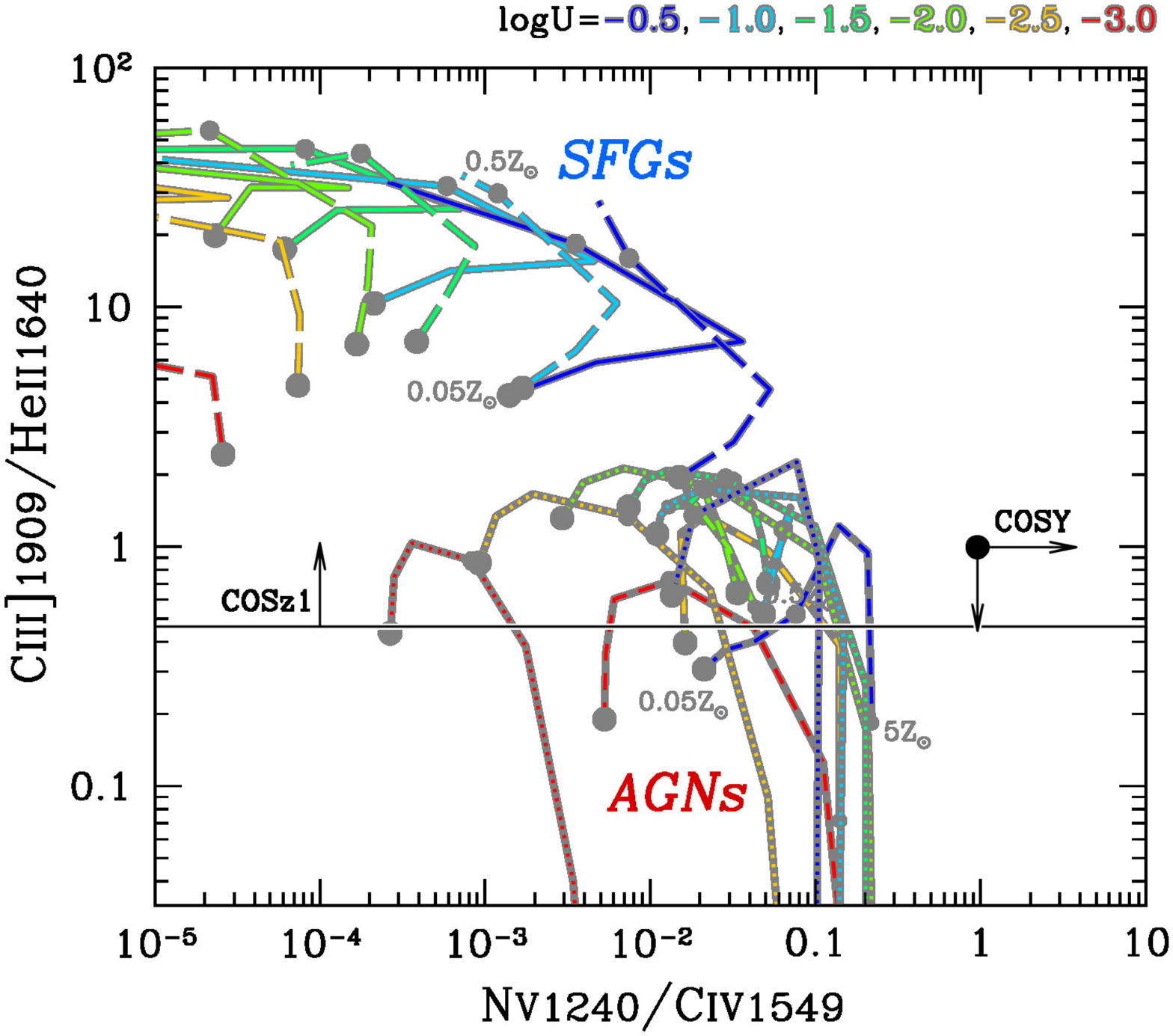} \\
\includegraphics[width=7cm, angle=-90]{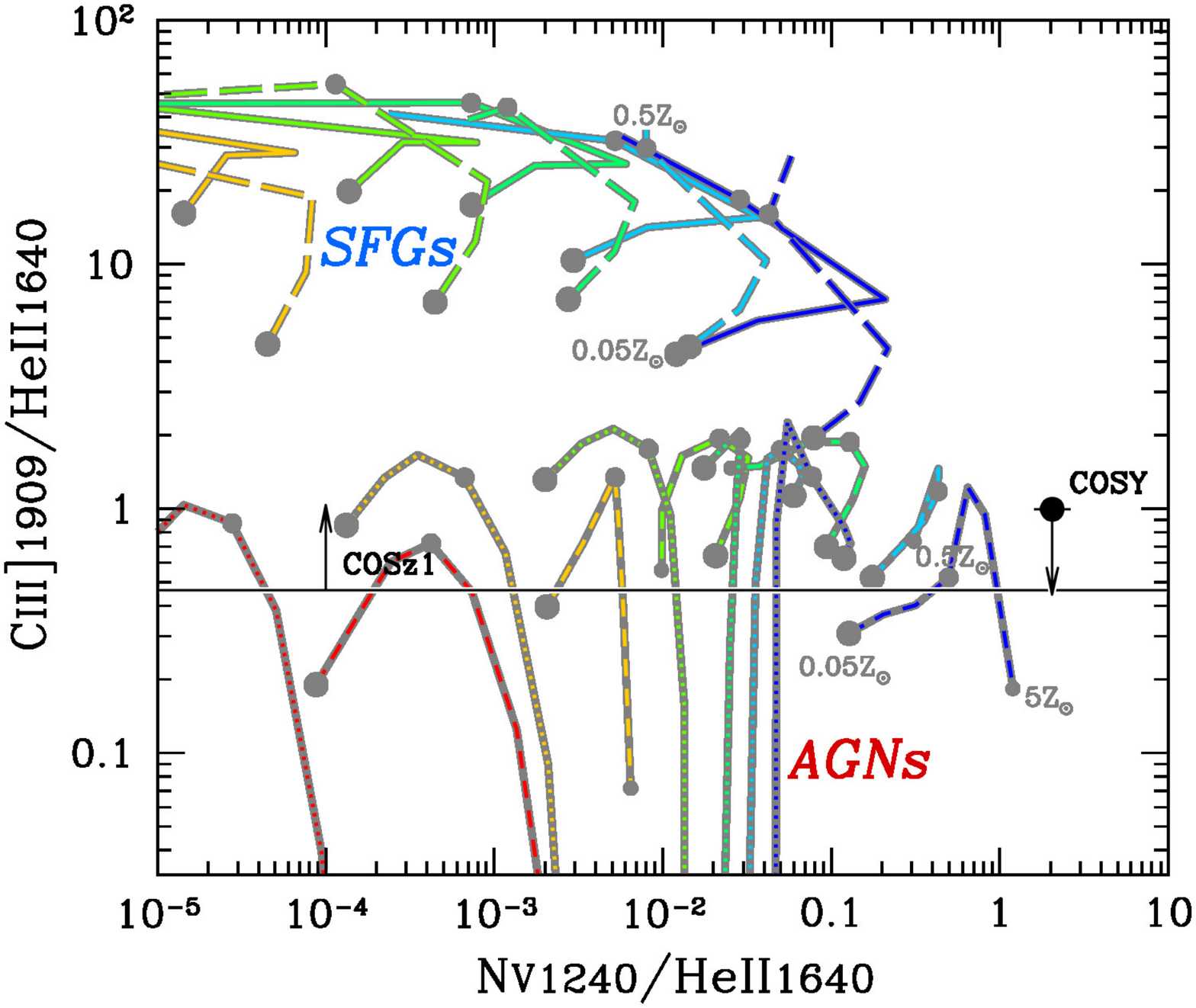}\\

\caption{\label{fig.models} 
Line ratio diagrams from photoionization models. Metallicity ranges from Z=0.05 to 1 (5) Z$_{\odot}$ for galaxy (AGN) models for an ionization parameter $\log(U)$ from -3.0 (red) to -0.5 (blue) as shown in the legend. Solid and long dashed lines are for single and binary stellar population models in star forming galaxies (SFGs), respectively. Dashed and dotted curves present AGN models with power-law indices $\alpha$=-1.2 (hard) and -2.0 (soft), respectively.
 }
\end{figure}

Figure \ref{fig.models} compares the CIII], CIV, He II and NV line ratios for COS$Y$ and COS$z1$ with the full range of models for star-forming galaxies of varying metallicity $Z$ and ionization parameter, $\log U$ and AGN models of varying power law indices $\alpha$. Since CIV$\lambda$1550.8\AA\ falls under a sky line, its contribution to the total CIV line flux is uncertain. However, regarding the typical ratio observed in low-$z$ galaxies between the two components,  any correction to include its contribution would not change our conclusions. In both figures, COSY is irreconcilable with a normal star-forming radiation field both on account of the low CIII]/He II ratio and, especially, the strength of NV. The prominence of NV is surprising and places it at the extreme end of the AGN predictions. For COS$z1$, the absence of He II and CIV provides a reasonable indication that the radiation field is consistent with a star-forming galaxy and perhaps the absence of Ly$\alpha$ is consistent with the source being atypical with its redder UV slope ($\beta=-1.18$) amongst the IRAC excess sample. Indeed, the detection of Ly$\alpha$ in COS$z2$ ($\beta=-1.72$) strengthens the case that IRAC excess sources have unusually powerful Ly$\alpha$ emission.

Our survey of three IRAC-excess $z\simeq$7 galaxies raises several points of interest. Firstly, it is clear how important it is to sample the full wavelength range where the key diagnostic lines occur. In the case of COS$z1$,  Figure \ref{fig.models} indicates how important are constraints on other diagnostic lines in reaching any firm conclusions. A second conclusion is the 
significant variation in the spectral properties of our trio of $z\simeq$7 targets. As Table \ref{tab.properties} reveals, all sources have similar luminosities and stellar masses yet the radiation field in COSY could not be more distinct from that in COS$z1$ and, probably COS$z2$ (noting the failure to provide a He II limit). Although this suggests the IRAC-excess sources may not represent a uniform population, if one includes the other 3 RBS sources, the detection of Ly$\alpha$ in 5 out of 6 such $z>6.8$ sources still supports the hypothesis that they lie in ionized bubbles. However, conceivably such ionized bubbles may be produced by more than one physical process as discussed by \citet{2017MNRAS.464..469S}.  COSY presents a convincing case for a AGN embedded in a star-forming galaxy whereas, perhaps others in the sample may lie in an over density of sources whose collective output is the primary cause of the ionized surroundings. Further imaging of IRAC-excess targets will help clarify this additional explanation.

Finally, our survey illustrates the feasibility of making further progress ahead of the launch of JWST. NIRSpec on JWST will transform studies of this kind by providing sensitive spectroscopic measures unhindered not only by the shielding effect of OH emission as in COS$z2$ but also the non-uniform atmospheric transmission. Such data will also, of course, finally ascertain the strength of the implied [O III] emission and, together with other rest-frame optical lines, thus provide a valuable independent constraint on the gas-phase metallicity, significantly improving the interpretation of diagrams such as Figure \ref{fig.models}.

\section{Summary}\label{sec:disc}

We have discussed the results from long exposure XSHOOTER spectra for three carefully-selected $z\simeq$7 targets with fairly uniform photometric properties,  each characterised by a prominent excess flux in one of the IRAC bands, suggestive of intense [O III] emission. We can summarize our finding as follows:

\begin{enumerate}

\item{} Our first goal has been to determine whether the ubiquity of Ly$\alpha$ emission in the four RBS IRAC-excess sources at $z>7$ is a distinct property of such sources, indicating each lies in an early ionized bubble. Ly$\alpha$ emission is now 
seen in one further source and, noting the range of UV continuum sloes, we conclude the hypothesis is still supported.

\item{} Our second goal has been to test whether a non-thermal radiation field is responsible for the putative ionized bubbles, as suggested by the early detection of CIII] and CIV in some of the RBS sources. We find a surprising diversity in the spectral characteristics of our IRAC-excess sources. One (COSY) shows evidence for a hard AGN component as evidenced by broad Ly$\alpha$ and prominent emission lines of He II and, especially, NV. The other two are consistent with normal star-forming galaxies although uncertainties remain and COSz2 also reveals broad Ly$\alpha$. This may suggest our IRAC-excess sources, which are amongst the most massive and luminous at these redshifts, are capable of ionizing their surroundings due to a variety of means, not only due to the presence of a AGN but also perhaps their location in an over density.
 
\item{} Finally, we have demonstrated the ultimate capability of ground-based spectrographs in this endeavour ahead of the launch of JWST. Our targets represent the brightest available in the HST archive, chosen carefully in a redshift range where most of the key diagnostic lines are visible. With integration times of over two nights with an efficient instrument, we have made some progress in constraining there nature of the radiation for sources well within the reionzation era.
\end{enumerate}

\acknowledgements{We acknowledge financial support from European Research Council Advanced Grant FP7/669253 (NL, RSE). DPS acknowledges support from the National Science Foundation through grant AST-1410155. It is a pleasure to thank Rychard Bouwens, Masami Ouchi, Roser Pell\'o and Daniel Schaerer for useful discussions. This work is based on observations taken by the CANDELS Multi-Cycle Treasury Program with the NASA/ESA HST, which is operated by the Association of Universities for Research in Astronomy, Inc., under NASA contract NAS5-26555. The CANDELS program also includes observations made with the Spitzer Space Telescope, which is operated by the Jet Propulsion Laboratory, California Institute of Technology under a contract with NASA. The work is based on data products from observations made with ESO Telescopes at the La Silla Paranal Observatory under ESO programmes ID 179.A-2005 and 097.A-0043 and data products produced by TERAPIX and the Cambridge Astronomy Survey Unit on behalf of the UltraVISTA consortium. Further data was taken with the W.M. Keck Observatory on Maunakea, Hawaii which is operated as a scientific partnership among the California Institute of Technology, the University of California and the National Aeronautics and Space Administration. This Observatory was made possible by the generous financial support of the W. M. Keck Foundation. The authors wish to recognize and acknowledge the very significant cultural role and reverence that the summit of Maunakea has always had within the indigenous Hawaiian community.  We are most fortunate to have the opportunity to conduct observations from this mountain. }

\bibliographystyle{aa}  
\bibliography{Laporte_AGN} 



\end{document}